# Nonlinear Photonic Neuromorphic Chips for Spiking Reinforcement Learning

Shuiying Xiang[1*], Yonghang Chen[1], Haowen Zhao[1], Shangxuan Shi[1], Xintao Zeng[1], Yahui Zhang[1], Xingxing Guo[1], Yanan Han[1], Ye Tian[1], Yuechun Shi[2], & Yue Hao[1]

**Photonic computing chips have made significant progress in accelerating linear computations, but nonlinear computations are usually implemented in the digital domain, which introduces additional system latency and power consumption, and hinders the implementation of fully-functional photonic neural network chips. Here, we propose and fabricate a 16-channel programmable incoherent photonic neuromorphic computing chip by co-designing a simplified Mach-Zehnder interferometer (MZI) mesh and distributed feedback lasers with saturable absorber (DFBs-SA) array using different materials, enabling implementation of both linear and nonlinear spike computations in the optical domain. Furthermore, previous studies mainly focused on supervised learning and simple image classification tasks. Here, we propose a photonic spiking reinforcement learning (RL) architecture for the first time, and develop a software-hardware collaborative training-inference framework (in-situ photonic training and hardware-aware fine-tuning) to address the challenge of training spiking RL models. We achieve large-scale, energy-efficient (photonic linear computation: 1.39 TOPS/W, photonic nonlinear computation: 987.65 GOPS/W) and low-latency (320 ps) end-to-end deployment of an entire layer of photonic spiking RL. Two RL benchmarks include the discrete CartPole task and the continuous Pendulum tasks are demonstrated experimentally based on spiking proximal policy optimization (PPO) algorithm. The hardware-software collaborative computing reward value converges to 200 (-250) for the CartPole (Pendulum) tasks, respectively, comparable to that of a traditional PPO algorithm. This experimental demonstration addresses the challenge of the absence of large-scale photonic nonlinear spike computation and spiking RL training difficulty, and presents a high-speed and low-latency photonic spiking RL solution with promising application prospects in fields such as real-time decision-making and control for robots and autonomous driving.**

[1] State Key Laboratory of Integrated Service Networks, State Key Discipline Laboratory of Wide Bandgap Semiconductor Technology, Xidian University, Xi'an 710071, China. [2]Yongjiang laboratory, No. 1792 Cihai South Road Ningbo, 315202, China. Correspondence and requests for materials should be addressed to S.Y. X. (email: syxiang@xidian.edu.cn )

The rapid development of artificial intelligence (AI), represented by large models and edge intelligence, has led to a continuous increase in the scale and performance of neural networks, generating a huge demand for computing capability, and making the issue of power consumption increasingly prominent. In contrast, the human brain can operate very complex and vast neural networks with a total power consumption of only 20 watts, which is significantly less than that of existing AI systems. Neuromorphic computing, which mimics the information processing mechanisms of neurons and synapses in the brain, offers a promising energy-efficient AI solution [1-4]. Traditional electronic computing chips struggle to meet the ever-increasing demands for computing power and energy efficiency in AI applications. Photonics computing, as a potential high-computing-power, low-power-consumption solution in the post-Moore era, has made significant progress in recent years [5-9], mainly including photonic multi-layer perceptrons [10-13], photonic convolutional neural networks [14-18], photonic diffractive neural networks [19-22], and photonic spiking neural networks (SNNs) [23-27]. The main photonic solutions include cascaded Mach-Zehnder interferometers (MZI) [28-37], micro-ring resonator (MRR) weight banks [38-43], cascaded Mach-Zehnder modulators (MZM) [44-45], phase change material (PCM) crossbars [46-47], and spiking laser neurons [48-53]. Reported photonic neural networks chips based on programmable MZI meshes have developed rapidly, which include matrix dimensions of $4 \times 4$ [10, 31], 6×6 [12, 30, 33], 8×8[13], 16×16 [34], 64×64 [36], 128×128 [37]. However, due to the inherently weak nonlinear effects of photons, most reported photonic computing chips can only perform linear computing acceleration in the optical domain, while nonlinear computing must still be carried out in the digital domain via opto-electronic (OE) and analog-to-digital (AD) conversions. The results are then sent back to the optical domain for further processing through digital-to-analog (DA) and electro-optical (EO) conversions, as indicated in Fig. 1(a). These frequent OE/EO and AD/DA conversions introduce additional system latency and increase power consumption. The lack of large-scale photonic nonlinear computation hinders the development of fully-functional photonic neural networks, which are essential for handling more complex and efficient computing tasks. Building a large-scale, fully-functional photonic neural network chip capable of performing both linear and nonlinear computations entirely in the optical domain remains an open problem.

Different from the traditional artificial neural networks (ANNs) that feature continuous-value activations and dense weight matrices, the SNNs employ binary spike activations and sparse weight matrices, as illustrated in Fig. 1(b). They possess an inherent low-power consumption advantage similar to the efficient information processing mechanisms of the human brain [54]. However, owing to the non-differentiability of spikes, the development of effective supervised training methods for SNNs has long been hindered. In recent years, ANN-to-SNN conversion methods and direct training approaches based on surrogate gradients have been proposed to enhance the performance of SNNs [55]. Photonic SNNs, which combine the advantages of photonic computing hardware and SNN architectures, offer an effective approach to achieving low-energy AI. Moreover, photonic SNNs alleviate the stringent requirements for high-precision AD/DA conversions when loading spike input and reading spike output results, representing a notably competitive technological paradigm for high-performance optical computing. However, previous studies of photonic SNNs mainly focused on supervised learning and simple image classification tasks.

Unlike traditional neural networks, which adopt a passive learning paradigm that relies on static, pre-defined datasets, reinforcement learning (RL) follows a dynamically active learning paradigm in which an agent interacts with the environment and learns the optimal behavior strategy according to reward feedback, playing a key role in autonomous decision-making and control in edge intelligence agents [56-58]. Among different RL algorithms, proximal policy optimization (PPO) algorithm has attracted lots of attention as it supports both discrete and continuous control tasks [59]. In particular, the spiking RL exhibits inherent energy-efficient advantage of SNN, which is critical for edge applications, but still faces significant training challenges [60-64]. Combining photonic neuromorphic computing with spiking RL is expected to build a more efficient, intelligent, and low power consumption computing system, further expanding the application scope of photonic computing to more complex scenarios. While photonic RL is still in its infancy [65-67], photonic spiking RL, in particular, has not yet been reported.

Here, we report the first demonstration of a large-scale, fully-functional (both linear and nonlinear), programmable incoherent photonic neuromorphic computing chip to deploy an entire layer of photonic spiking RL with 272 trainable model parameters. Through three key advances in increasing chip scale via different materials, proposing photonic spiking RL architecture, and developing software-hardware collaborative training algorithms, we realize low-latency, high-energy-efficiency and high-computing-density spike-based photonic linear and nonlinear computations.

1. In terms of chip design, a simplified 16×16 MZI mesh tailored for SNNs was designed for implementing photonic linear matrix computation, and a 16-channel distributed feedback laser with saturable absorber (DFB-SA) array was optimized for implementing element-wise nonlinear spiking activation. For the first time, the optical-domain nonlinear spiking neuron array is scaled up to 16 channels, and the trainable model weights and activation threshold parameters for photonic SNN are increased to 272 in total (16×16+16). This approach addresses the challenge of the lack of large-scale optical-domain nonlinear spike computation and lays the foundation for the development of fully-functional photonic SNN chips.
2. On the architectural aspect, we proposed a photonic spiking RL architecture for the first time. We deployed a spiking PPO algorithm on the photonic neuromorphic chip, which uses a spike-based actor network and a continuous-value-based critic network. The actor network is tested for inference directly on the photonic neuromorphic chip. This opens a novel path for realizing autonomous decision-making and control through photonic neuromorphic computing.
3. To address the challenges of photonic SNNs on training, we proposed a software-hardware collaborative training-inference framework, consisting of global software pre-training, local hardware in-situ training and hardware-aware software fine-tuning, which effectively enhance inference accuracy and mitigate the impact of analog computing noise and manufacturing process variations. Two reinforcement learning benchmarks including the discrete CartPole task and the continuous Pendulum task are demonstrated experimentally, which provides a prototype for photonic spiking RL algorithm.

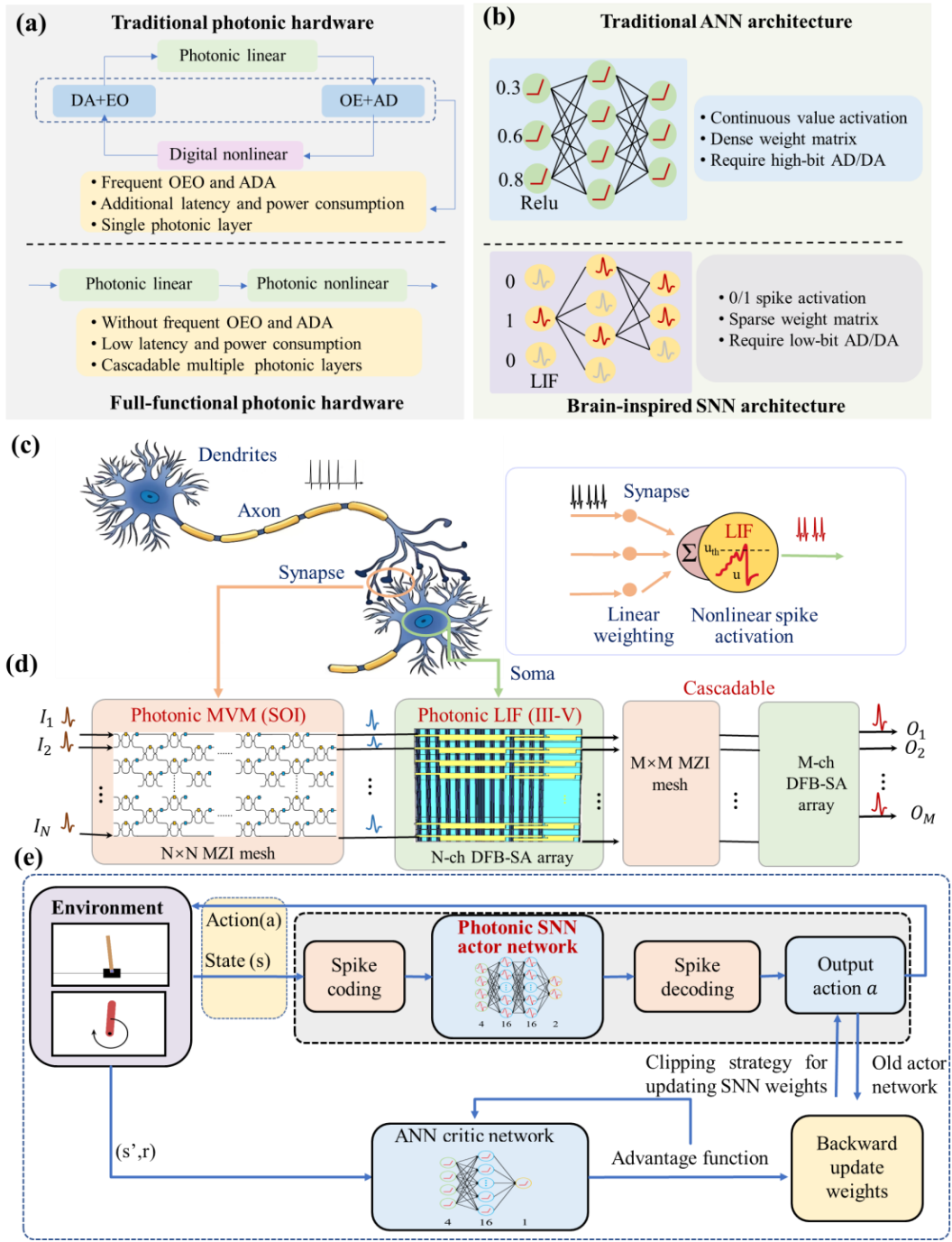

Fig. 1. Concept of incoherent photonic neuromorphic chips. (a)Challenges of traditional photonic hardware and characteristics of fully-functional photonic hardware, (b) challenges of traditional ANN and characteristics of brain-inspired SNN, (c) schematic diagram of biological neuron and LIF neuron, (d) schematic diagram of photonic MVM based on silicon-on-insulator (SOI) material and photonic LIF neuron based on III-V material, (e) the overall architecture of photonic spiking RL based on PPO algorithm.

In our experimental demonstration, the incoherent photonic neuromorphic computing chip achieves energy efficiency and computing density of 1.39 TOPS/W and 0.13 TOPS/mm$^2$ for photonic linear computation, and 987.65 GOPS/W and 533.33 GOPS/mm$^2$ for photonic nonlinear computation, respectively. The end-to-end computing latency of an entire photonic SNN layer is 320 ps. For the testing state-action pairs, the accuracy of hardware inference decreased by only 1.5% (2%) compared to the pure software algorithm for the discrete (continuous) task, and the hardware-software collaborative computing reward values converge to 200 (-250) for the CartPole (Pendulum) tasks. Our demonstrated programmable incoherent photonics SNN chip and end-to-end software-hardware collaborative training-inference framework can be generalized to various photonic SNN architectures, and will be promising for wide resource-constrained applications such as edge computing, autonomous driving and embodied intelligence.

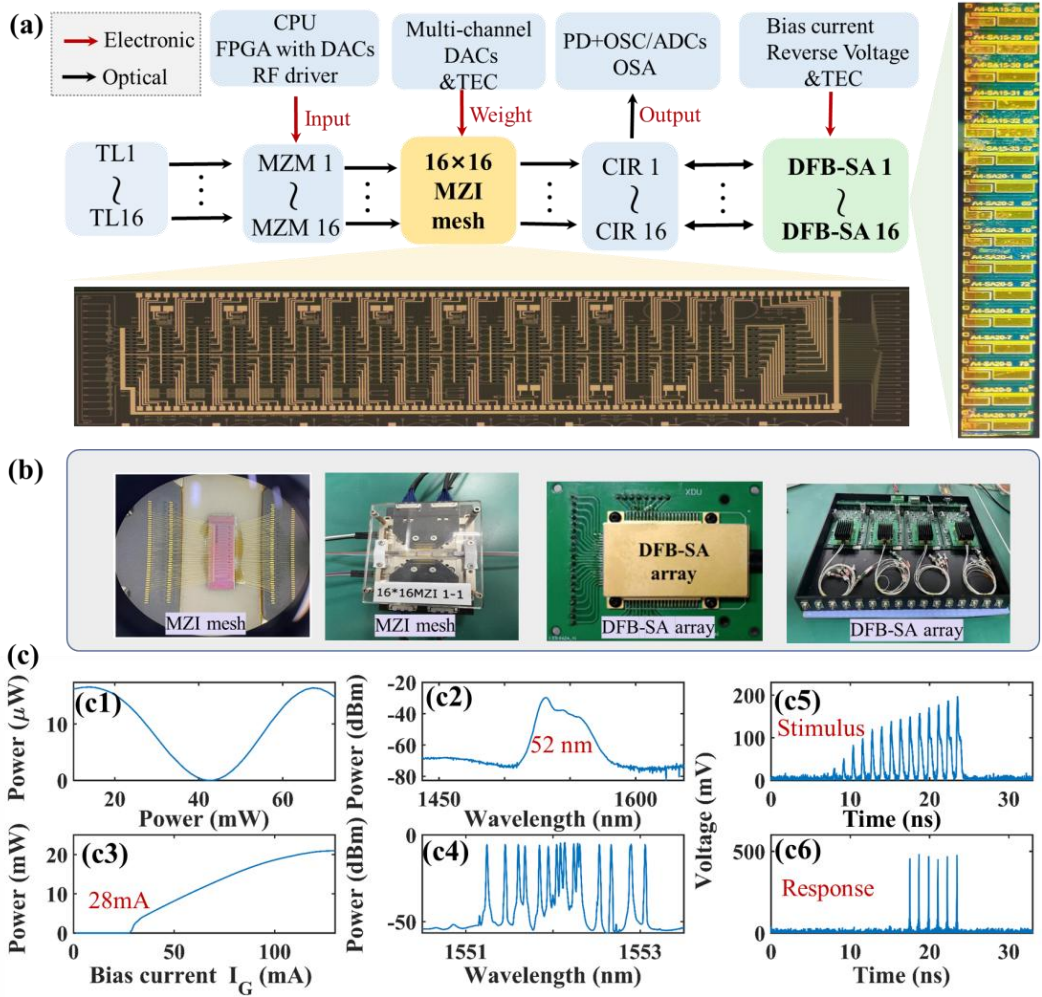

Fig. 2. Schematic diagram of experimental setup for testing the photonic neuromorphic computing chip modules and the measured chip characteristics. (a) Experimental setup and microscopic images of the fabricated MZI mesh chip and DFB-SA array chips. (b) Photos of MZI chip modules and DFB-SA array compact packaged modules, (c) characteristics of the MZI mesh and DFB-SA laser, (c1) transfer function and (c2) optical spectrum of MZI chip, (c3) output power as a function of the bias current of DFB-SA laser, (c4) optical spectrum of DFB-SA laser subject to incoherent optical injection, (c5) stimulus and (c6) temporal response of DFB-SA laser, the temperature is fixed to 25°C, gain current $I_G$ =32 mA, and reverse voltage $V_{SA}$ = -1.8 V.

**Results**
**Concept and architecture.**

As presented in Fig. 1(a), unlike most photonic hardware designed for accelerating linear computation, the fully-functional photonic hardware enables the implementation of both linear and nonlinear computation in the optical domain. Figure 1(b) illustrates the main difference between traditional ANNs and brain-inspired SNNs. The binary spike activation property of an SNN results in a sparse weight matrix and requires only low-bit AD/DA to load inputs and read outputs, making it hardware-friendly.

Biological neurons are the fundamental working units of the nervous system. A typical biological neuron includes dendrites, soma, axon, and synapse, as shown in Fig. 1(c). The dendrites are the input part of the neuron, responsible for receiving and integrating signals from other neurons. The soma integrates signals from the dendrites and decides whether to generate an action potential, i.e., a spike. The axon functions as the output pathway, propagating spike signals to downstream neurons. Synapses connect different neurons and are responsible for transmitting information. The strength of synapses can change, forming the biological basis for learning and memory, known as synaptic plasticity. These components work together to enable neurons to receive, process, and transmit information, thereby supporting the complex functions of the brain and nervous system.

Drawing on the mechanisms of brain information processing, the core units of brain-like SNNs are synapses that perform linear weighting operations and spiking neurons that perform nonlinear spike activation operations. Among these, the leaky integrate-and-fire (LIF) neuron is a widely used spiking neuron model in SNNs due to its simplicity. The output of a spiking neuron is the result of the linearly weighted summation of the output of all the neurons in the previous layer, followed by a nonlinear spike activation. Specifically, an entire layer of an SNN includes matrix-vector multiplication (MVM) and LIF spike activation functions.

**Fully-functional incoherent photonic neuromorphic chip.**

In our work, both linear MVM and nonlinear LIF spike activation functions are demonstrated with photonics chips in the optical field, as presented in Fig. 1(d). In each layer, signals propagate through a linear MVM operation, followed by element-wise nonlinear spike activation. We consider different materials to achieve large-scale high-performance photonic MVM and LIF chips. On the one hand, a photonic synapse array chip based on a simplified MZI mesh was designed and fabricated on a silicon photonic platform to perform incoherent optical linear MVM. On the other hand, a 16-channel DFB-SA laser array was implemented on a III-V platform to enable element-wise nonlinear spike activation. Thus, the fabricated incoherent photonic neuromorphic chip is capable of implementing an entire layer of hardware SNN in the optical domain.

**Photonic spiking RL architecture based on PPO.**

We propose a photonic spiking RL architecture based on PPO, as shown in Fig. 1e. The RL agent consists of an actor network based on a SNN and a critic network based on an ANN. The SNN actor network is used to generate the probability distribution of actions according to the current state. During inference, the SNN actor network is deployed onto the photonic neuromorphic chip. The ANN critic network is used to estimate the value of the current state. By utilizing the collected trajectory data, the value of the advantage function for each state is calculated. The advantage function represents the degree of advantage of a certain action relative to the average policy in the current state, that is, the degree to which taking this action can obtain a higher reward than the average reward. We consider two standard RL tasks from OpenAI Gym: (1) CartPole: a discrete control task where the agent stabilizes a pole in the upright position by moving a cart left or right. (2) Pendulum: a continuous control task where the agent applies torque to quickly stabilize a pendulum in the upright position. The state spaces, the action spaces, and reward functions for both tasks are detailed in Supplementary Section 1.

During the spiking RL process, the agent interacts with the environment to obtain a series of information such as states, actions, rewards, and the next state. The collected data is used to calculate the value of the advantage function for each state. Then, the PPO strategy is used to update the SNN actor network. The reward information in the collected trajectory data and the value estimation of the next state are used to update the ANN critic network.

The photonic neuromorphic chips for deploying the SNN actor network include the photonic linear computing chip based on the MZI array and the photonic nonlinear computing chip based on the DFB-SA array. To match the scale of the photonic neuromorphic chip, we optimize the scale of the SNN actor network for the discrete task to be 4×16×16×2, with the ANN critic network size being 4×16×1. While for the continuous task, the scale of the SNN actor network is 3×16×16×1, and the scale of the ANN critic network is 3×32×32×1.

**Chip characterization and testing.**

We design and construct a testing system to deploy the spiking RL on photonic neuromorphic chips. The schematic diagram of the experimental setup, along with the fabricated MZI mesh chip and 16-channel DFB-SA laser array chip, are presented in Fig. 2(a). Detail descriptions can be found in the Methods section. In this system, incoherent optical input is generated from 16 tunable lasers (TLs) operating at different wavelengths. This incoherent network requires only direct detection instead of coherent detection and is not sensitive to the phase instability of the input light. The input vector is generated by the FPGA board and modulates the light intensity via the MZM array, and the weight matrix is loaded on the phase shifters of the MZI mesh chip. The optical output of MZI mesh chip is the linear weighting result of MVM, which is then unidirectionally injected into the 16-channel DFB-SA array via a 16-channel three-port optical circulator array for nonlinear spiking activation. The compactly packaged modules of the MZI mesh and the DFB-SA array are depicted in Fig. 2(b).

**Photonic MVM based on MZI mesh chip tailored for SNN.**

Conventional MZI meshes based on singular value decomposition for complex-valued MVM requires $2N^2$ phase shifters with a depth of $(2N+1)$ and adopts coherent detection. This results in high chip loss, tuning complexity, and sensitivity to phase instability. Note that, SNNs typically have sparse weight matrices and exhibit high tolerance to weight errors. Leveraging this property, we simplify the photonic synapse array chip design by employing an $(N+1) \times (N+1)$ unitary MZI mesh followed by an $N \times N$ diagonal matrix to approximate the functionality of an $N \times N$ synaptic weight matrix in the proposed simplified MZI mesh [68], resulting in a total of $N \times (N+1)/2 + N$ phase shifters. Each MZI contains only a single phase shifter on one of its inner arms. All phase shifters are thermally tuned by TiN heaters. Thus, to represent a $16 \times 16$ weight matrix, the design uses $16 \times (16+1)/2 + 16 = 152$ phase shifters, with a network depth reduced to $N + 2$. The overall structure of the simplified MZI mesh for a $16 \times 16$ weight matrix is detailed in Supplementary Section 2.

Figure 2(c1) presents the transfer curve of the phase shifter on the MZI chip, where the transfer curve of the first port is tested. The result shows that as the phase shifter voltage V changes, the corresponding loaded power ($V^2/R$) varies accordingly, and the output power follows a sinusoidal dependence on the loaded power, with a half-wave phase shift power $P_\pi$=30 mW. Figure 2(c2) displays the corresponding transmission spectrum of the MZI mesh chip, with a 3-dB bandwidth of approximately 52 nm. To determine loss, the input light is injected into one input port, and the powers from all 16 output ports are detected and summed. The measured loss for the packaged MZI mesh chip is about 13 dB. This simplified design tailored for SNN offers promising advantages of small footprint (2.37 mm×8.25 mm= 19.55 mm$^2$) and low-loss (13 dB).

**Photonic LIF based on DFB-SA array chip.**

The DFB-SA laser includes a gain region and a saturable absorber (SA) region. The gain region is driven by a current source, denoted as the gain current $I_G$, while the SA region is reversely driven by a voltage source, denoted as $V_{SA}$. As shown in Fig. 2(c3), the measured lasing threshold is about 28 mA and is almost the same for all the 16 lasers. The epitaxial wafer structure and a scanning electron microscope (SEM) picture for the DFB-SA laser unit are presented in Supplementary Section 3, alogh with the self-pulsation outputs and the corresponding frequency spectra and optical spectra for different $I_G$ and $V_{SA}$, as well as the neuron-like responses. The maximum self-pulsation frequency is about 5 GHz. In this work, we optimized the epitaxial design of the DFB-SA laser to achieve higher self-pulsation frequency (from 2 GHz to 5 GHz), lower lasing threshold (from 86 mA to 28 mA) and smaller chip area (single unit area from 0.45 $mm^2$ to 0.075 $mm^2$) [26]. Moreover, the channel count has been scaled to 16 for the first time.

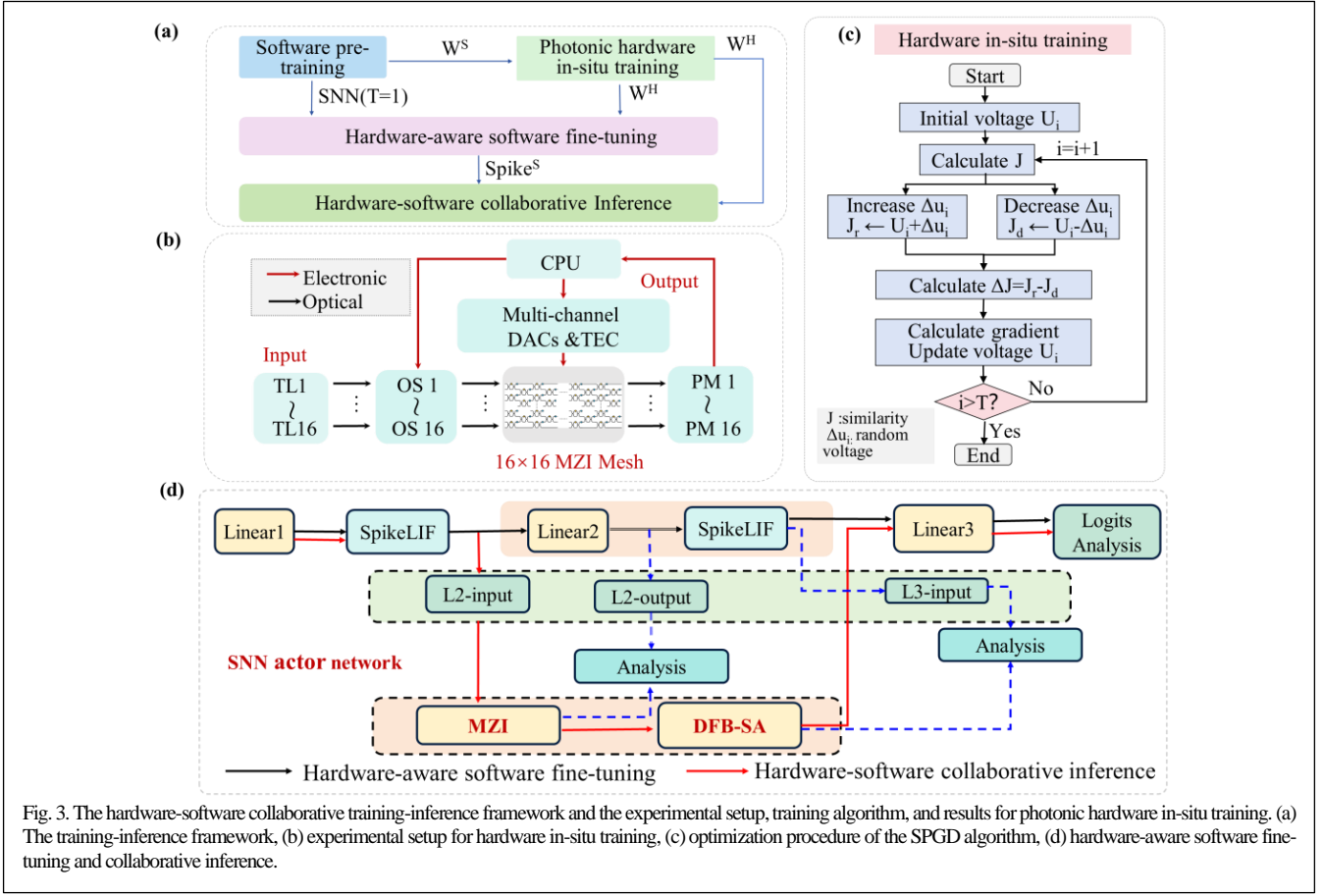

Fig. 3. The hardware-software collaborative training-inference framework and the experimental setup, training algorithm, and results for photonic hardware in-situ training. (a) The training-inference framework, (b) experimental setup for hardware in-situ training, (c) optimization procedure of the SPGD algorithm, (d) hardware-aware software fine-tuning and collaborative inference.

Figure 2(c4) presents the optical spectrum of the DFB-SA laser subjected to incoherent multi-wavelength optical injection when the DFB-SA laser operates as a photonic spiking neuron. The neuron-like nonlinear threshold response is also presented. Figure 2(c5) shows the stimulus signal input to the DFB-SA laser, and Fig. 2(c6) shows the nonlinear threshold response output of the DFB-SA laser. The results indicate that for low input amplitude, the DFB-SA laser does not generate spike outputs. However, when the input amplitude exceeds the activation threshold, the DFB-SA laser can achieve a threshold response similar to that of a biological neuron, with the activated spike being independent of the input amplitude.

**Software-hardware collaborative training-inference framework.**

Hardware photonic computing inevitably suffers from accuracy degradation due to fabrication errors and system noise. Many in-situ or adaptive training methods for photonic neural network chips have been reported [33, 69-72]. To address both the training challenges of SNN and the accuracy loss in optical computing, we propose an end-to-end software-hardware (software pre-training, local photonic hardware in-situ training, and hardware-aware software fine-tuning) collaborative training-inference framework for photonics SNNs. The entire hardware and software training and inference stages are shown in Fig. 3(a).

Step 1: Software pre-training stage. The actor network adopts aN SNN, and the critic network adopts an ANN. The SNN is trained from scratch using the surrogate gradient method with single time step (T = 1).

Step 2: Photonic hardware in-situ training of MZI chip. The 16×16 weight matrix of the hidden-layer linear transformation in the SNN actor network trained from Step 1 is exported for online training on the MZI mesh chip. To match the photonic SNN chip, the exported weights are clipped to move negative weights and biases terms. After the MZI training is completed, the resulting weight matrix is imported back into the algorithm, with the middle linear layer weights fixed and its gradients disabled during retraining. The hardware setup for in-situ training is shown in Fig. 3(b). Here, we adopt the stochastic parallel gradient descent (SPGD) algorithm to map the trained weights to the phase-shifter heater voltages of the MZI mesh chip [33]. The SPGD optimization procedure is presented in Fig. 3(c). For more detailed description on the experimental setup and SPGD algorithm of the hardware in-situ training, please refer to the Methods section.

Step 3: Hardware-aware software fine-tuning. With the weights fixed, the network is re-trained in a hardware-aware manner to compensate for the accuracy loss caused by fabrication imperfections and system noise. After the training is completed, three datasets-the input to the middle linear layer (L2-input), its output (L2-output), and the input to the LIF neurons (L3-input)-are exported for subsequent collaborative inference, as presented in Fig.

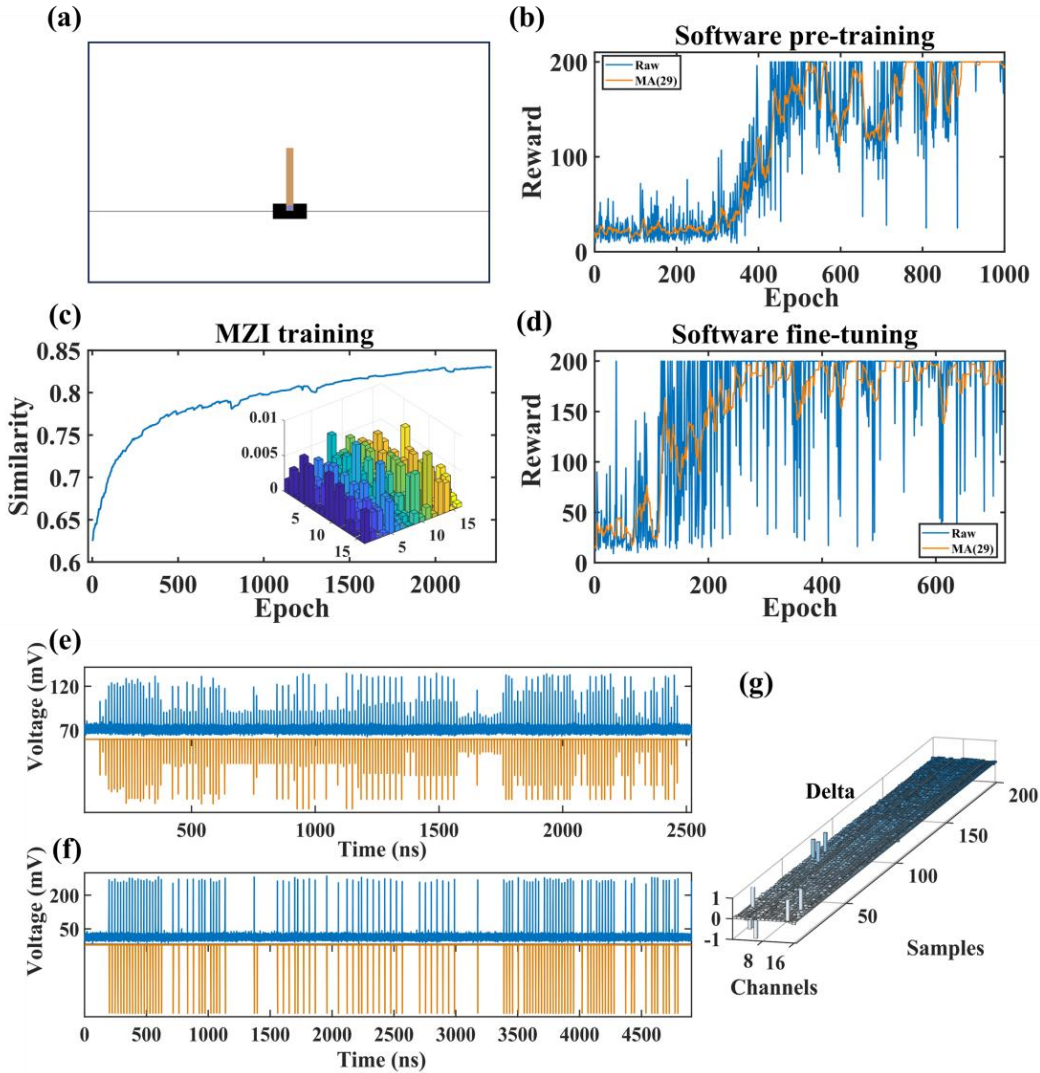

Fig. 4. The results for the training and inference stage of the discrete task. (a) CartPole task, (b) reward for software pre-training, (c) MZI training for discrete task, (d) hardware-aware fine-tuning trained reward, (e) experimentally measured linear and (f) nonlinear outputs for discrete task, (g) error of state-action sample pairs.

3(d). Note, the environment is configured for a maximum of 200 execution steps per training episode. Thus, the exported data corresponds to this upper limit.

Step 4: Hardware-aware collaborative inference. The hardware training results are returned to the algorithm for hardware-aware collaborative inference. The output results of the DFB-SA array are returned to the algorithm and passed through the third linear layer. The obtained results are compared item by item with those from the fine-tuned software model to calculate the hardware-software collaborative computation accuracy.

**Discrete control task using photonic neuromorphic chips.**
We first consider a discrete control task implemented with the spiking PPO algorithm and the photonic neuromorphic computing chips. The environmental and training parameters used in the algorithm are provided in Supplementary Section 4.

Figure 4(a) shows the test environment interface for the CartPole discrete task. Figure 4(b) presents the reward curve for the baseline software algorithm. The blue curve represents the raw data, while the orange curve denotes the moving-window average with a window size of 29. The results indicate that the reward value first reached 200 in the 429-th training epoch and gradually converged to a stable value after the 886-th epoch. Figure 4(c) displays the hardware training results when the weight of the hidden-layer linear transformation are deployed onto the MZI mesh chip. The similarity between the algorithm-trained weights and those mapped onto the MZI chip is calculated to access the fidelity of weight mapping. Here, the formula for calculating the similarity between matrix A and matrix B is: $a = flatten(A)$, $b = flatten(B)$, $cf = \dfrac{a \cdot b}{\|a\|\|b\|}$.

The results show that the similarity approaches 0.80 after approximately 1000 training epochs and finally reaches 0.83 after 2300 epochs. Figure 4(d) shows the reward curve during the hardware-aware software fine–tuning stage. Here, an early-stopping strategy is adopted, when the performance does not improve for 200 consecutive rounds starting from the 722-nd epoch, the training is stopped. The reward value first reaches 200 in the 38-th training epoch and gradually converges to a stable value after the 258-th epoch with slight fluctuations. Obviously, the convergence speed in the fine-tuning stage is significantly faster and more stable than that in the baseline software training.

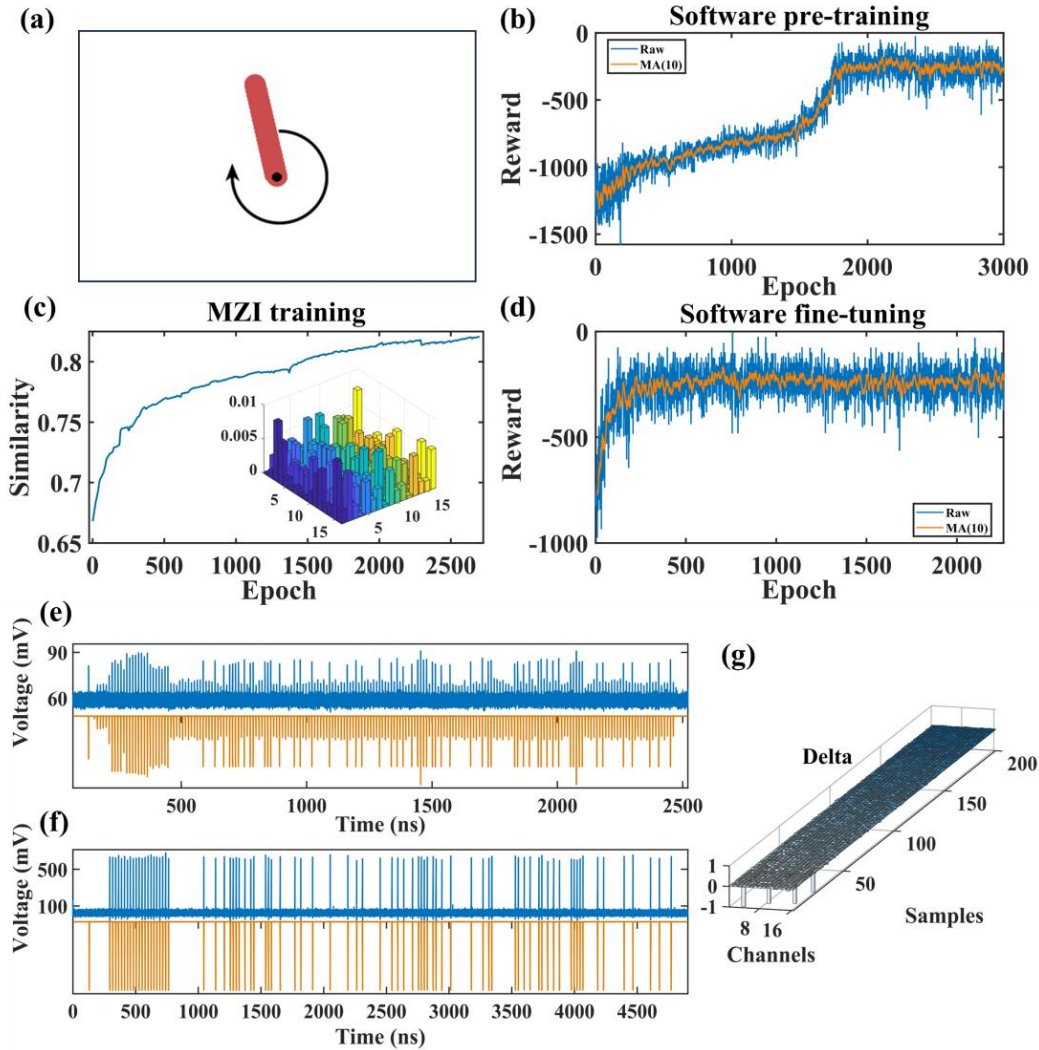

Fig. 5. The results for the training and inference stage of the continuous task. (a) Pendulum task, (b) reward for software pre-training, (c) MZI training for continuous task, (d) hardware-aware fine-tuning trained reward, (e) experimental measured linear and (f)nonlinear outputs for continuous task, (g) error of state-action sample pairs.

In the hardware inference experiment, we considered 200 state-action sample pairs due to the memory limit of the FPGA board. Among them, the Linear2 (L2, $16 \times 16$) hidden layer was deployed on the photonic neuromorphic chip. The input data of the L2 are generated through the FPGA, and the modulators were used to transform the 16-channel inputs of L2 into 16 optical signals, which were then fed into the MZI array to perform linear MVM.

Figure 4(e) shows the MZI linear computing results from channel 1, where the orange curve represents the ideal target values. The results indicate that the actual MZI outputs closely match the algorithm-calculated targets. Figure 4(f) shows the nonlinear activation outputs of the DFB-SA array measured in the experiment compared with the target activation results obtained in the algorithm. It can be found that the DFB-SA was not activated at three time instants: 1395.99 ns, 3692.93 ns, and 4114.82 ns. Figure 4(g) shows the difference distribution between experimental and ideal spike activations, where values near 1 indicate excessive excitation (6 occurrences) and values near -1 indicate insufficient excitation (16 occurrences). This corresponds to an error rate of 0.6875% at the current layer, yielding a final-layer inference accuracy of 98.5%.

**Continuous control task using photonic neuromorphic chips.**
In addition to the discrete task discussed above, the spiking PPO algorithm can also be used to solve a continuous control task. Here, we further evaluate the photonic spiking RL algorithm by implementing the Pendulum task. The environmental and training parameters used in the algorithm are provided in Supplementary Section 5.

Figure 5(a) shows the test environment interface for the Pendulum continuous task. Figure 5(b) displays the reward value curve obtained by the baseline software algorithm. The blue curve represents the raw data, and the orange curve represents the moving-window average with a window size of 10. The results indicate that the reward value reaches -204.65 at the 1750-th training epoch, and stabilizes at around -250 with minor fluctuations thereafter. Figure 5(c) shows the results of the hardware online training of the MZI chip.

The similarity approaches 0.80 after approximately 1500 epochs and finally reaches 0.82 after 2700 epochs. Figure 5(d) shows the results of the hardware-aware fine-tuning stage. Here, the early-stopping strategy is also applied, when the performance does not improve for 1500 consecutive epochs starting from the 2258-th epoch, the training is stopped. The reward value can reach -127.59 at around the 71-st training epoch, and stabilizes at around -230 with minor fluctuations in the subsequent training process. Similar to the discrete task, the hardware-aware fine-tuning stage converges faster and yields a more stable reward value than the baseline software training.

In the photonic hardware inference experiment, we also considered 200 state-action pairs samples. Here, we also deployed L2 of the $16\times16$ hidden layer on the photonic neuromorphic chip. Figure 5(e) shows the MZI linear computing results for channel 1 and the target algorithm outputs, indicating a close match. Figure 5(f) shows the nonlinear spike activation outputs of the DFB-SA array compared with the algorithm results. Compared to the target output, the DFB-SA was not activated at 130.325 ns, and one spike was less emitted. Figure 5(g) shows the difference distribution: the number of values equal to 1 is zero, and the number equal to -1 is four. This corresponds to an error rate of 0.125% at the current layer, yielding a final-layer inference accuracy of 98%.

## Discussion

Here, we further explore the performance for larger MZI chip scale. In 2025, reported MZI array chips have reached scales of $64\times64$ [36] and $128\times128$ [37], both supporting positive and negative weights. Therefore, in our algorithm, we also considered SNN actor networks with hidden layers of $16\times16$, $64\times64$, and $128\times128$, and remove the constraints of non-negative weights. The hidden layer scale of the critic network is considered the same as that of the corresponding actor network. The results are presented in Table 1. For the discrete task, the converged epoch refers to the first training epoch at which the reward reaches 200, while for the continuous task, it denotes the epoch at which the training curve begins to stabilize. The corresponding training curves are provided in Supplementary Section 6. The ablation experiments show that the network with larger hidden-layer scales converges faster than those with a $16\times16$ hidden layer.

We also compare the reward performance with different time steps T for both tasks. Here, we consider T=1, T=2, and T=4. As shown in Table 1, a larger T leads to faster convergence. Detailed results are provided in Supplementary Section 6.

Tabel 1. The converged epoch for different network scale and T.

| Ablation study | Parameter | Converged epoch (CartPole) | Converged epoch (Pendulum) |
|---|---|---|---|
| Scale | 16×16 | 84 | 1445 |
| | 64×64 | 98 | 874 |
| | 128×128 | 69 | 579 |
| T | 1 | 429 | 1763 |
| | 2 | 66 | 994 |
| | 4 | 61 | 946 |

**Metrics.**

For both the MZI mesh chip and the 16-channel DFB-SA laser array chip, we evaluate the metrics for photonic linear and nonlinear computations. The calculation methods of these metrics are provided in Supplementary Section 7. For the MZI mesh chip, the throughput of the linear MVM operation can be estimated as 2.5 TOPS, where we consider a 5 GHz clock frequency limited by the maximum self-pulsation frequency of the DFB-SA laser. The chip area is 19.55 mm$^2$, and the total power consumption is about 1.8 W. Therefore, the energy efficiency is 1.39 TOPS/W, and the computing density is 0.13 TOPS/ mm$^2$. For the DFB-SA laser array chip, the throughput can be calculated as 640 GOPS. The chip area of the 16-channel DFB-SA array is 1.2 mm$^2$. The energy consumption, including the gain and SA regions, is approximately 0.648 W. Thus, the energy efficiency is 987.65 GOPS/W, and the computing density is 533.33 GOPS/mm$^2$. The latency for implementing an entire layer of photonic SNN, i.e., including photonic MVM and photonic LIF neuron, is calculated as 320 ps. We also compare the performance with the state-of-the-art photonic neural network chips in Table 2. To the best of our knowledge, among photonic neural network chips supporting optical nonlinear computation, our chip achieves the largest number of trainable parameters.

**Scalability.**

The photonic spiking neurons based on the DFB-SA laser array can be readily scaled up to 150 channels with high wavelength precision [73]. The MZI chip can also be scaled up to 128×128 with acceptable loss [37]. Moreover, it is practical to fabricate compact hybrid-integrated photonic neuromorphic chips consisting of photonic spiking neuron chiplets and photonics synapse chiplets through advanced packaging and heterogeneous integration technologies.

## Conclusions

We have demonstrated a programmable incoherent photonic neuromorphic computing chip with 272 trainable model parameters for deploying an entire layer of spiking RL, which can perform both linear and nonlinear spike computations entirely in the optical domain. The photonic synapse chip, based on a simplified 16×16 MZI mesh tailored for SNN, offers advantages of low loss, low power consumption and small footprint. The photonic spiking neuron array chip, based on an optimized 16-channel DFB-SA array, achieves a low lasing threshold and compact size through an improved epitaxial wafer structure. We further develop a software-hardware collaborative training-inference framework, comprising software pre-training, photonic hardware in-situ training, and hardware-aware software fine-tuning, to enhance the accuracy of the photonic SNNs. This framework can be generalized to any photonic SNN hardware architecture. We propose a photonic spiking RL architecture for the first time, which is composed of an SNN actor network and an ANN critic network. During inference, the spiking actor network generates actions based on the state and is deployed onto the photonic neuromorphic chip. Through software-hardware collaborative optimization, efficient collaborative computing between the spiking actor network of the spiking PPO and the photonic neuromorphic chip has been achieved. For the CartPole discrete task, the reward of spiking PPO converges stably to 200, comparable to that of a traditional ANN-PPO algorithm, fully verifying its effectiveness for discrete action tasks. For the Pendulum continuous action space task, the reward converges to -250, demonstrating that the photonic spiking RL architecture is capable of solving complex tasks in continuous action spaces. After deploying the spiking actor network onto the photonic neuromorphic chip, 200 pairs of state-action pairs were selected for testing, and the accuracy reached 98%.

Table 2. Comparison with state-of-the-art photonic neural network chips that support optical nonlinear computation.

| Metrics | Computing density | Energy efficiency | Trainable parameters | Optical nonlinear computation | For SNN |
|---|---|---|---|---|---|
| **Our work** | 0.13 TOPS/mm$^2$ | 1.39 TOPS/W | **272** | √ | √ |
| Ashtiani et al [11] | 1.75 TOPS/mm$^2$ | 2.9 TOPS/W | 67 | √ | X |
| Bandyopadhyay et al [12] | 0.02 TOPS/mm$^2$ | 0.013 TOPS/W | 132 | √ | X |
| Feldmann et al [23] | N/A | N/A | 64 | √ | √ |

Compared with the state-of-the-art photonic neural network chips that support optical nonlinear computation, our chip exhibits the advantages of high energy efficiency (1.39 TOPS/W for linear and 987.65 GOPS/W for nonlinear computations in optical domain) and high computing density (0.13 TOPS/mm$^2$ for linear and 533.33 GOPS/mm$^2$ for nonlinear computations in optical domain), as well as low latency (320 ps). Our work addresses the absence of large-scale photonic nonlinear spike computation and the training challenges of spiking RLs, providing a high-speed and low-latency photonic spiking RL architecture. It is expected to play an important role in many fields such as real-time robotic decision-making and autonomous driving, and to promote the cross-integration and development of photonic computing chips, SNNs and deep RL.

## Methods

### Device fabrication.

**MZI mesh chip**. The MZI mesh chips are manufactured using 248 nm deep ultraviolet (DUV) lithography and inductively coupled plasma (ICP) etching on a standard 220 nm silicon-on-insulator (SOI) wafer. In the packaging part, the SOI chip is glued to a printed circuit board (PCB), and the metal wires are connected to the PCB via wire-bonding method. When the light is injected from a single input port, the measured chip loss is about 13 dB.

**DFB-SA laser array chip**. The DFB-SA lasers are fabricated using a P-I-N diode structure grown on the AlGaInAs/InP material system. The total length of the cavity is $L_{cavity}=300$ μm, the length of SA region is $L_{SA}=10$ μm, the width of the laser chip is 250 μm, and the ridge waveguide width is 2.5 μm. Additionally, anti-reflection and high-reflection (HR) coating are applied to the two laser facets to enhance the light emission power, and the SA region is positioned near the HR side. In order to achieve appropriate saturable absorption, we optimized and designed parameters such as the number of quantum wells in the laser. The optimized multi-quantum well structure contains 7 well layers and the thickness of the upper and lower confined layers are around 70 nm. The grating are fabricated with the reconstruction-equivalent-chirp technique, which allows for a large-scale laser array with high wavelength precision [73].

### Measurement setup.

The experimental setup for testing the photonic neuromorphic computing prototypical system is illustrated in Fig. 2 (a). The input and weight signals were generated using an FPGA board (Zynq UltraScale+ RFSoC ZCU216) equipped with a high-speed DA array (9.85GHz,14bit). The FPGA was controlled by a digital computer. To generate optical carriers, 16-channel continuous-wave (CW) TLs were utilized. The input signals were modulated using the MZM. Polarization controllers were employed to align the polarization state. The modulated outputs of the MZMs were then optically injected into the MZI mesh chip, where the weights were mapped to the phase shifter voltages of the MZI mesh chip via 12-bit multi-channel DACs. The output of the MZI mesh chip is then unidirectionally injected into the 16-channel DFB-SA laser array via three-port optical circulators array. The outputs of the 16 DFB-SAs were respectively detected by a photodetector (PD, Agilent/HP 11982A) and then recorded by an oscilloscope (OSC, Keysight DSOZ592A).

### Local photonic hardware in-situ training.

The optical matrix represented by the MZI mesh is implicit, meaning there is no one-to-one mapping between the phase shifter heater voltages and the matrix elements. Here, the SPGD algorithm for the local in-situ training is adopted to configure the MZI mesh chip to represent the target weight matrix $W^S$. However, due to noise, fabrication errors, and the limited precision of the multi-channel DACs that drive the phase shifters, the trained hardware weights $W^H$ may not perfectly match $W^S$.

As shown in Fig. 3(b), the setup for configuring the MZI mesh weights includes a 16-channel CW TL sources, 16-channel optical switches (OS), multi-channel voltage sources, 16-channel optical power meters (PM), a thermoelectric cooler (TEC), and a host computer. The host computer controls the multi-channel optical switches through a serial port to sequentially inject light from each input port of the MZI mesh, and collects the output powers from all 16 output ports through the power meters, thereby forming the hardware weight matrix. The phase-shifter voltages are updated by the SPGD algorithm, with the host computer loading the voltages onto the MZI mesh via the multi-channel voltage sources over Ethernet. The temperature of the MZI chip is stabilized at 25°C by the TEC.

During the photonic hardware in-situ training process, we firstly load a randomly initialized voltage $U_i$ for each phase shifter of MZI. For these 152 voltages, we can measure a set of optical power values from 16-channel MZI output ports when the CW light is injected from one MZI input port. These 16 optical power values are regarded as one column of weights. By sequentially injecting CW light from each input port, we can obtain 16 columns of weights, yielding the complete hardware weight. The similarity J between the hardware weight and target weight [33] is then computed by flattening both matrices into vectors and calculating their cosine similarity. Next, we consider a random perturbation voltage $\Delta \mathbf{u_i}$, and the positive and negative perturbations $U_i+\Delta \mathbf{u}_i$ and $U_i-\Delta \mathbf{u}_i$ are then loaded onto the phase shifter voltages, respectively. Correspondingly, we can repeat the measurements mentioned above, and get $J_r$ for positive perturbation and $J_d$ for negative perturbation. After that, the similarity change is calculated as $\Delta J = J_r - J_d$. The gradient with respect to $\Delta \mathbf{u}_i$ is then estimated from $\Delta J$, and the phase-shifter voltages are updated accordingly. By repeating the process iteratively, the hardware weight can be gradually converged to the target weight.

**Data availability**. The data that support the findings of this study are available from the corresponding author upon reasonable request.

**Code availability.**
The source codes of this study are available in the repository of the project.

**Acknowledgements**

This work was supported by the National Key Research and Development Program of China (2021YFB2801900); National Natural Science Foundation of China (No. 61974177, 62125503, 62261160388); The Fundamental Research Funds for the Central Universities (QTZX23041).

**Author contributions**

S. Y. Xiang designed the overall architecture and experiments. S. X. Shi optimized the SNN algorithm. H. W. Zhao, Y. H. Zhang, X. X. Guo performed experimental measurements. Y. C. Shi designed and fabricated the DFB-SA array chip. Y. Tian designed and fabricated the MZI mesh chip. S. Y. Xiang prepared the manuscript. S. Y. Xiang., and Y. Hao directed all the research. All authors analyzed the results and implications and commented on the manuscript at all stages.

**Competing interests**

The authors declare no competing interests.

**Supplementary information**

Supplementary information is available for this paper.